# Optical properties of graphane in infrared range

E. I. Preobrazhensky*, I. V. Oladyshkin, M. D. Tokman
*Institute of Applied Physics of the Russian Academy of Sciences*
*evgenypr@ipfran.ru

**Abstract**

The theory of optical effects in hydrogenated graphene (graphane) in terahertz and infrared range is developed, including the analysis of complex conductivity, reflection coefficient for graphane on a substrate and dispersion of surface plasmon-polaritons. The calculations are based on quite simple analytical approximation of graphane band structure in the vicinity of Γ-point and on the modified model of quantum coherence relaxation. Comparison of the obtained theoretical results with corresponding experimental data can be used both for the determination of graphane characteristics (Fermi level, relaxation rate etc.) and for the investigation of potential applications of this material in the design of new optical elements.

**I Introduction**

In recent years, two-dimensional materials closely related to pure graphene [1], such as bilayer and multilayer graphene and graphane (hydrogenated graphene) [2–10], have become the subject of intense theoretical and experimental study. The main distinguishing features of these materials are associated with the formation of a finite bandgap instead of the zero gap (i.e., the intersection of the valence and conduction bands at Dirac points). In this work we study graphane, which has a unique property of controllable adsorption and desorption of hydrogen atoms (for example, by annealing), including the complete return of the monolayer to the pure graphene state. With gradual hydrogenation, it is possible to obtain 2D materials with some specific parameters, for example, electrical or thermal conductivity. In this regard, graphane is considered as a promising material for the fabrication of hydrogen cells [11], the production of biosensors [12], transistors [13], and also elements for spintronics [14].

The valence band of graphane consists of two subbands [15, 16] which can be attributed to "heavy" and "light" holes (see [17, 18]). In Ref. [16] these subbands near the Γ-point were considered and the averaged mass of charge carriers in the valence band was estimated, i.e. the effective mass of electron-hole excitons. Obviously, the average mass approximation is inapplicable for calculation of conductivity at frequencies comparable to the characteristic frequency of intersubband transitions. Another problem for correct description of graphane electromagnetic properties in this frequency range is relatively short time of quantum coherence relaxation in graphene-based materials. Thus, in graphene it can be as small as several femtoseconds [19-21]. Under these conditions the standard model for the relaxation of the density matrix off-diagonal terms, or so called "τ-approximation" [22-24][1], leads to a number of fundamental incorrectness in determining the electrodynamic response in the terahertz and far-infrared ranges. First of all, the continuity equation is violated if the perturbations of current and electric charge are calculated independently [33]. The correct expressions for the conductivity (or permittivity) of a fermion ensemble in a 2D crystal lattice in the case of finite dissipation are given in Ref. [25]. In particular, it was demonstrated in Ref. [25] that the theoretical prediction for the permittivity of graphene in far-IR range significantly differs when passing from the standard to the modified relaxation operator. A number of important nonlinear and quantum optical effects in graphene and topological isolators are closely interconnected with 2D plasmon-polariton modes [1, 26-31].

---
[1] This model usually leads to the replacement $\omega \Rightarrow \omega + i\tau^{-1}$ when calculating a linear electrical conductivity $\sigma(\omega)$.

The main aim of the present paper is the theoretical study of graphane optical properties in the range of intersubband resonant frequencies. Comparison of the theoretically predicted and properly measured reflection, transmission and absorption coefficients in line with the dispersion of surface plasmons could directly verify the basic models of graphane band structure. For example, the experimental data on the propagation length of surface plasmons would make it possible to find the relaxation time of intersubband and intrasubband quantum coherence in graphane in the terahertz and infrared frequency ranges.

The paper is organized as follows. In Chapter II the state functions of the system are calculated in the framework of model Hamiltonian proposed in Ref. [16] (for the subbands of light and heavy holes near the Γ-point). In Chapter III we generalize the theoretical results from Ref. [25] to the case of graphane and obtain the expressions for the permittivity in the frequency range where the influence of intersubband transitions is fundamentally important. In Chapter IV we give the expressions for the coefficients of electromagnetic radiation reflection and transmission through the graphane on a dielectric substrate and for the 2D plasmon dispersion. Finally, Chapter V is devoted to the results of numerical calculations.

Notice that during the hydrogenation of a double-layer or multilayer graphene we can expect the appearance of two separated (but not touchnig) subbands in the valence band. In this regard in Appendix we analyze numerically a 2D system with the modified Hamiltonian which contains the energy gap between two subbands as a parameter.

**II. Model of graphane valence band**

Let us consider a graphane monolayer lying in the plane *xy*. In paper [16] in the framework of invariant method [32] the following model ($\boldsymbol{k} \cdot \boldsymbol{p}$)-Hamiltonian was proposed for the vicinity of Γ-point in the valence band:

$$\widehat{H}_{\boldsymbol{k}}^{(v)} = -a\hat{I}(k_x^2 + k_y^2) - b[\sigma_z(k_x^2 - k_y^2) + 2\sigma_x k_x k_y], \tag{1}$$

and $\widehat{H}_{\boldsymbol{k}}^{(c)} = \hbar^2 \hat{I} \frac{k_x^2 + k_y^2}{2m_c}$ for the conduction band. Here $\boldsymbol{k} = \boldsymbol{x_0} k_x + \boldsymbol{y_0} k_y$, where $\hbar \boldsymbol{k}$ is the electron quasi-momentum (in the plane of monolayer), $\boldsymbol{x_0}$ and $\boldsymbol{y_0}$ are unit vectors, $a > b > 0$, $\sigma_{x,z}$ are standard Pauli matrixes, $\hat{I}$ is the identity matrix, $m_c$ is the effective mass of charge carriers in the conduction band. The following dispersion law for the energies of heavy and light holes $E_\pm(k^2)$ corresponds to the Hamiltonian (1) [16]:

$$E_\pm = -(a \mp b)k^2, \tag{2}$$

where $k = \sqrt{k_x^2 + k_y^2}$. Notice that two curves $E_\pm(k^2)$ touch each other at the point $k = 0$.

In the conduction band, separated from the valence band by the energy gap $W_{vc}$, we have a common dependence $E = W_{vc} + \hbar^2 k^2 / 2m_c$. Thus, the described model gives one branch of the energy dispersion in the conduction band and two branches in the valence bands, touching each other at the point $k = 0$ (see Fig. 1).

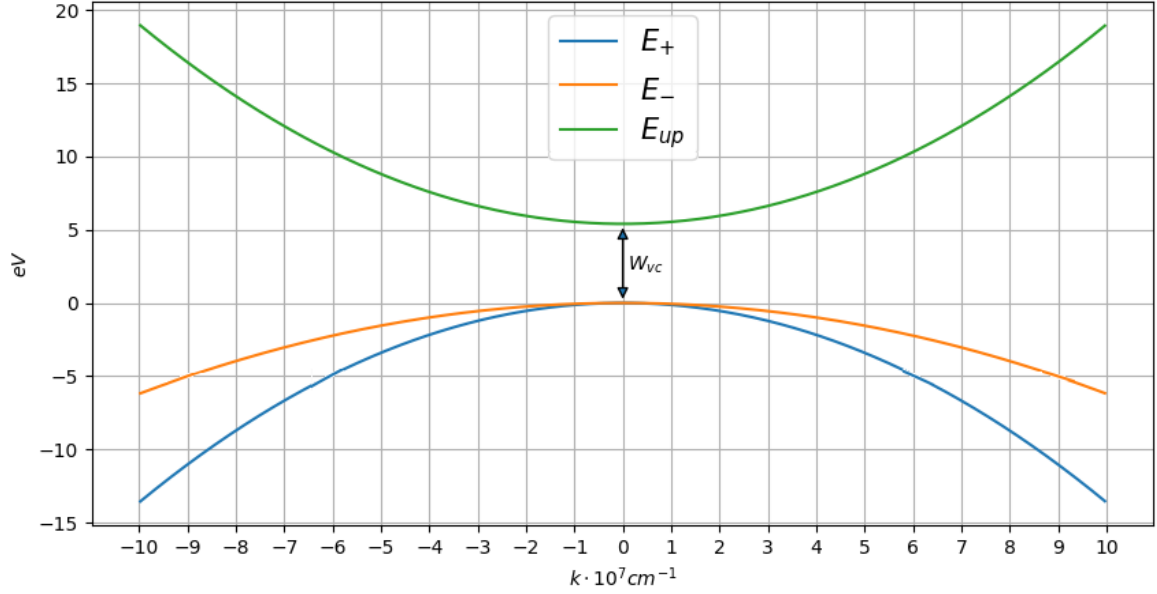

**Fig. 1.** Model band structure of graphane in the vicinity of Γ-point [16] $W_{vc} = 5.4\ eV, a = -\frac{1.31}{m_0}, b = -\frac{0.49}{m_0}$, $m_0$ is the free electron mass in vacuum.

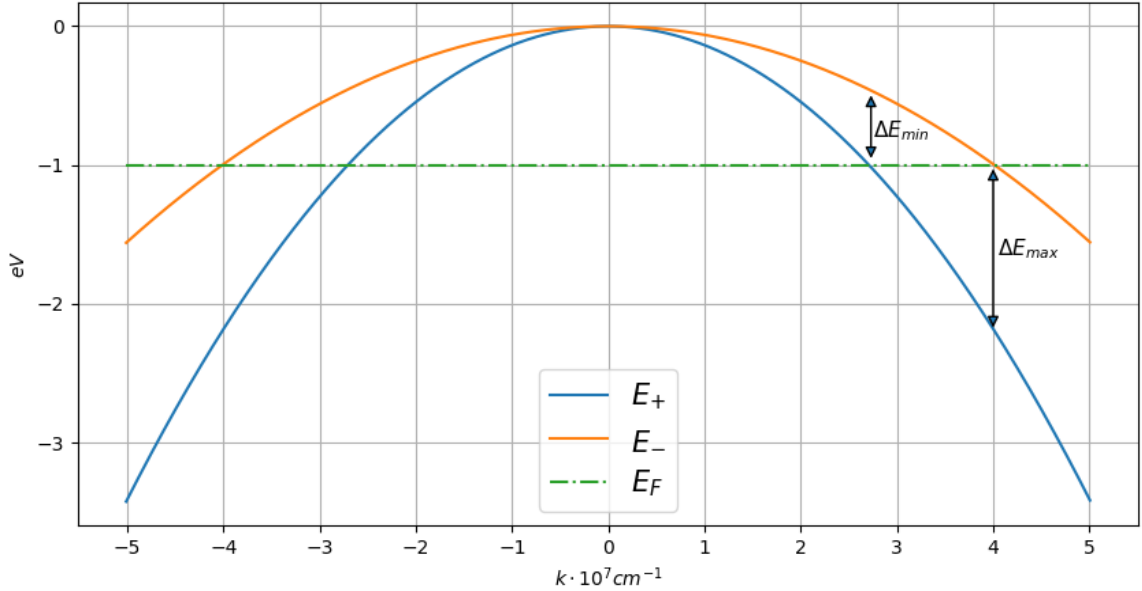

**Fig. 2.** The Subbands of light and heavy holes in the vicinity of Γ-point [16]. Dash-dotted line shows the Fermi level, two arrows show possible direct intersubband transitions with maximal and minimal energies ($E_F = -1\ eV, a = -\frac{1.31}{m_0}, b = -\frac{0.49}{m_0}$).

Considering the response of the graphane monolayer to the action of electromagnetic field at frequency $\omega$ when $\hbar\omega \ll W_{vc}$ and $E_F < 0$ (where $E_F$ is the Fermi energy), we can take into account only charge carriers in the valence band. Here we use the Hamiltonian (1) corresponding to the band structure shown in Fig. 2, so that the state vector has the form $\Psi = \begin{pmatrix} u_{1k} \\ u_{2k} \end{pmatrix} e^{-ikr}$. Using the Schrödinger equation

$$\left(\widehat{H}_k^{(v)} - \widehat{I}E\right)\begin{pmatrix}u_{1k}\\u_{2k}\end{pmatrix} = 0,$$

we obtain the expression for the state vector in **k**-representation:

$$\begin{pmatrix}u_{1k}\\u_{2k}\end{pmatrix}_{(+)} = \begin{pmatrix}\cos\theta_k\\\sin\theta_k\end{pmatrix}, \quad \begin{pmatrix}u_{1k}\\u_{2k}\end{pmatrix}_{(-)} = \begin{pmatrix}-\sin\theta_k\\\cos\theta_k\end{pmatrix}, \quad (3)$$

where $\theta_k$ is the angle between the *x*-axis and the vector **k**.

Let us find the expression for the matrix element for a dipole moment of direct transition between the states with the energies $E_{(+)}(k^2)$ and $E_{(-)}(k^2)$. Since the system is isotropic in the plane of monolayer, it suffices to find the matrix element of the *x* coordinate. In **k**-representation, we can use the following relation:

$$x_{(+)\boldsymbol{k}(-)\boldsymbol{k}} = \begin{pmatrix}u_{1k}\\u_{2k}\end{pmatrix}_{(+)}^* \frac{i\partial}{\partial k_x}\begin{pmatrix}u_{1k}\\u_{2k}\end{pmatrix}_{(-)},$$

whence it follows that

$$\left|x_{(+)\boldsymbol{k}(-)\boldsymbol{k}}\right|^2 = \frac{\sin^2\theta_k}{k^2}. \quad (4)$$

Notice that the obtained expression for $\left|x_{(+)\boldsymbol{k}(-)\boldsymbol{k}}\right|^2$ differs from the corresponding expression for interband transitions in graphene near the Dirac point only by a numerical coefficient equal to 4 (see, for example, Ref. [25]).

### III. Conductivity of graphane monolayer

To calculate the surface conductivity of a 2D dissipative system with a given band structure we will use the results from Ref. [25]. In this paper a modified operator of density matrix relaxation was proposed, so that it does not violate the continuity equation for the current density and charge density perturbations in a dissipative system. Assuming that the characteristic electron wavenumber $\langle k\rangle \sim \sqrt{\frac{|E_F|}{a+b}}$ is large compared to the inverse spatial scale of the electromagnetic field inhomogeneity $L^{-1}$, we can use the homogeneous field approximation[2]. In this case, we obtain the following expressions for the surface conductivity from the formulas given in [25]:

$$\sigma(\omega) = \sigma_{inter}(\omega) + \sigma_{intra}(\omega) \quad (5)$$

$$\sigma_{intra}(\omega) = \frac{e^2 g}{4\pi^2}\sum_{\alpha}\iint_{\infty}\frac{\left(\eta\frac{\partial}{\partial k}E_{\alpha k}\right)\left(\eta\frac{\partial}{\partial k}f_{\alpha k}\right)}{\hbar^2(i\omega - 2\gamma_{\alpha k\alpha k})}d^2k \quad (6)$$

$$\sigma_{inter}(\omega) = \frac{e^2 g}{4\pi^2}i\omega\sum_{\alpha\neq\beta}\iint_{\infty}\frac{|\eta\cdot r_{\alpha k\beta k}|^2(f_{\alpha k}-f_{\beta k})(E_{\alpha k}-E_{\beta k})}{(E_{\alpha k}-E_{\beta k})^2 - \hbar^2\omega^2 - 2i\hbar^2\omega\gamma_{\alpha k\beta k}}d^2k \quad (7)$$

Here we separated the conductivity components due to intraband ($\sigma_{intra}$) and interband ($\sigma_{inter}$) transitions, respectively; also here $g$ is the degeneracy factor, $e$ is the elementary charge; $\alpha$ and $\beta$ are band indices, $E_{\alpha k}$ and $f_{\alpha k}$ are the energy and the occupation probability for the state $|\alpha k\rangle$ respectively, $r_{\alpha k\beta k} = \langle\alpha k|r|\beta k\rangle$ is the matrix element of a particle coordinate in the plane of the monolayer for "direct" transitions, $\eta$ is the unit vector parallel to the electric field projection on the plane of graphane. The values $\gamma_{\alpha k\beta k} = \gamma_{\beta k\alpha k}$ and $\gamma_{\alpha k\alpha k}$ are intraband and interband relaxation constants for quantum coherence, respectively. In order to avoid misunderstandings, we note that the value $\gamma_{\alpha k\alpha k}$ is defined as the inverse relaxation time of off-diagonal matrix elements $\rho_{\alpha k\alpha k'}$ in the limit $|\boldsymbol{k}-\boldsymbol{k}'|\ll|\boldsymbol{k}|$, i.e. this is the quasi-momentum relaxation time. This value in principle may significantly differ from the inverse

---

[2] for the interband component of conductivity this approximation corresponds to the model of "direct" transitions

relaxation time of diagonal elements of the density matrix $\rho_{\alpha k \alpha k} \equiv f_{\alpha k}$, which corresponds to the energy relaxation time. In the limit of zero dissipation, Eqs. (5-7) exactly correspond to the standard Kubo formula (see the proof in Ref. [25]).

Considering an isotropic system in the plane of the layer, we can choose $\boldsymbol{\eta} = \boldsymbol{x_0}$ in Eqs. (5-7) without loss of generality. Keeping in mind the presence of two subbands in the valence band, further we use the following notations in Eqs. (5-7): $\alpha, \beta$ are denoted as "$\pm$", $\gamma_{(+)k(-)k} = \gamma_{(-)k(+)k} \approx \gamma_{inter}$ and $\gamma_{(+)k(+)k} \approx \gamma_{(-)k(-)k} \approx \gamma_{intra}$. For a degenerate Fermi-Dirac distribution we obtain:

$$\sigma_{intra}(\omega) = i \frac{e^2 g}{4\pi \hbar^2 \omega} \frac{1}{1+i\frac{2\gamma_{intra}}{\omega}} \sum_{\pm} \left| k \frac{\partial E_{\pm}(k)}{\partial k} \right|_{E_{\pm}=E_F} \tag{8}$$

$$\sigma_{inter}(\omega) = -i\omega \frac{e^2 g}{2\pi^2} \int_{\Delta E_{min}}^{\Delta E_{max}} \left( \int_0^{2\pi} |x_{(+)k(-)k}|^2 d\theta_k \right) \frac{k \Delta E \left| \frac{d\Delta E}{dk} \right|^{-1} d\Delta E}{(\Delta E)^2 - \hbar^2(\omega^2 + 2i\omega \gamma_{inter})} \tag{9}$$

where $\Delta E(k) = E_{(+)}(k) - E_{(-)}(k)$, $E_F$ is the Fermi level, the values $\Delta E_{max,min}$ are the points of intersection of the Fermi level with the dependencies $E_{\pm}(k)$ (see **Fig. 2**).

The expressions (8) and (9) lead to fairly general results for the case of power dependence $|E_{\pm}(k)| \propto k^{n_{\pm}}$. First, for the intraband conductivity we obtain:

$$\sigma_{intra}(\omega) = i \frac{e^2 g}{4\pi \hbar^2 \omega} \frac{|E_F| \sum_{\pm} n_{\pm}}{1+i\frac{2\gamma_{intra}}{\omega}}. \tag{10}$$

Second, if $|\Delta E(k)| \propto k^n$, then for the standard dependence[3] of the coordinate matrix element on the quasimomentum

$$\int_0^{2\pi} |x_{(+)k(-)k}|^2 d\theta_k = A k^{-2}$$

we obtain

$$\sigma_{inter}(\omega) = -i \frac{e^2 g}{4\pi^2 \hbar} \frac{A n^{-1}}{\sqrt{1+2i\gamma_{inter}/\omega}} \ln \left( \frac{\Delta E_{max} - \hbar\omega\sqrt{1+2i\gamma_{inter}/\omega}}{\Delta E_{max} + \hbar\omega\sqrt{1+2i\gamma_{inter}/\omega}} \times \frac{\Delta E_{min} + \hbar\omega\sqrt{1+2i\gamma_{inter}/\omega}}{\Delta E_{min} - \hbar\omega\sqrt{1+2i\gamma_{inter}/\omega}} \right), \tag{11}$$

where $A$ is an arbitrary constant. In particular, when $\gamma_{inter}/\omega \to 0$ from Eq. (11) we get:

$$\text{Re}\, \sigma_{inter}(\omega) = \frac{e^2 g A n^{-1}}{4\pi \hbar} \left[ \Theta\left(\omega - \frac{\Delta E_{min}}{\hbar}\right) - \Theta\left(\omega - \frac{\Delta E_{max}}{\hbar}\right) \right], \tag{12}$$

where $\Theta(x)$ is the Heaviside step function. Note that in a 2D quantum well with several parabolic subbands ($n = 2$), the absorption dependence on frequency has the form of a set of plateaus similar to one described by Eq. (12), which was observed experimentally in the paper [18].

In further consideration of graphane electromagnetic properties in IR region, we use the expressions (10) and (11) to determine its conductivity. The valence band is set as two touching parabolas, which is presupposed by the Hamiltonian (1). Therefore, we use:
a) for the calculation of intraband conductivity: $\sum_{\pm} n_{\pm} = 4$.
b) for the calculation of interband conductivity [4]: $A = \pi$, $n = 2$, $\Delta E_{min} = \frac{2b|E_F|}{a+b}$, $\Delta E_{max} = \frac{2b|E_F|}{a-b}$.

---

[3] which follows, for example, from dimensional analysis
[4] for comparison, in case of a graphene $A = \frac{\pi}{4}$, $n = 1$ and $\Delta E_{max} \to \infty$ in the vicinity of the Dirac point.

## IV. Optical properties of the monolayer

Let us consider a monolayer with the surface conductivity $\sigma(\omega)$ lying on a substrate with the permittivity $\epsilon$.

### A. Transmission and reflection coefficients

Let a plane electromagnetic wave be incident normally to the sample. Then the standard solution for the reflection (*r*) and transmission (*t*) coefficients for the complex amplitude of the electromagnetic field gives:

$$r = \frac{\sqrt{\epsilon}-1+\frac{4\pi}{c}\sigma(\omega)}{\sqrt{\epsilon}+1+\frac{4\pi}{c}\sigma(\omega)}, \quad t = \frac{2\sqrt{\epsilon}}{\sqrt{\epsilon}+1+\frac{4\pi}{c}\sigma(\omega)} \tag{13}$$

and for power reflection and transmission coefficients we get $R = |r|^2$, $T = \frac{1}{\sqrt{\epsilon}}|t|^2$.

### B. 2D surface plasmon-polaritons in graphane

Now let us move to surface electromagnetic waves propagating in the plane of the monolayer (here we consider propagation along the *x*-axis, see **Fig. 3**).

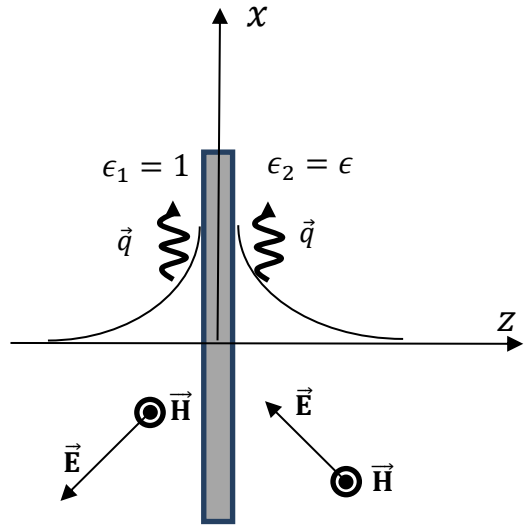

Fig.3. A surface plasmon-polariton propagating along a monolayer lying on a substrate with permittivity $\epsilon$ ($z > 0$).

Dispersion equation for surface plasmon-polaritons (SPP) for this configuration of electromagnetic field is given in Ref. [26]:

$$\frac{4\pi i \sigma(\omega)}{\omega} + \frac{\epsilon}{\sqrt{q^2 - \epsilon \frac{\omega^2}{c^2}}} + \frac{1}{\sqrt{q^2 - \frac{\omega^2}{c^2}}} = 0, \tag{14}$$

where $q$ is the wavenumber of the surface mode, the values $\sqrt{q^2 - \epsilon \frac{\omega^2}{c^2}}$ and $\sqrt{q^2 - \frac{\omega^2}{c^2}}$ describe the inverse lengths of the electromagnetic field decay in the half-space filled by the substrate and vacuum, respectively. In the limit $q^2 \gg \epsilon \frac{\omega^2}{c^2}$, the Eq. (14) corresponds to the quasi-electrostatic approximation for the surface plasmon and so

$$\frac{4\pi i \sigma(\omega)}{\omega} q + \epsilon + 1 = 0. \tag{15}$$

In the case of negligible interband conductivity and dissipation, from the latter expression it follows that $\omega \propto \sqrt{|E_F|q}$, which coincides with the surface plasmon dispersion in graphene [1] and topological insulators [26].

## V. Numerical calculations

A. *Surface conductivity*

In this section, the result of numerical calculations of the functions $\text{Re}[\sigma(\omega)]$ and $\text{Im}[\sigma(\omega)]$ for different relaxation rates $\gamma$ and Fermi levels $E_f$ are presented.

Consider the frequency dependence of the surface conductivity of graphane in the case of different relaxation rates $\gamma$. Here we choose relatively large Fermi level ($-0.5\ eV$) to emphasize the key features of the conductivity related to the formation of the plateau mentioned in Section III (see the frequency dependence of the real component of $\sigma(\omega)$). For more detailed analysis, we give the graphs for the intersubband (Fig. 4.) and intrasubband (Fig. 5.) parts of the surface conductivity separately.

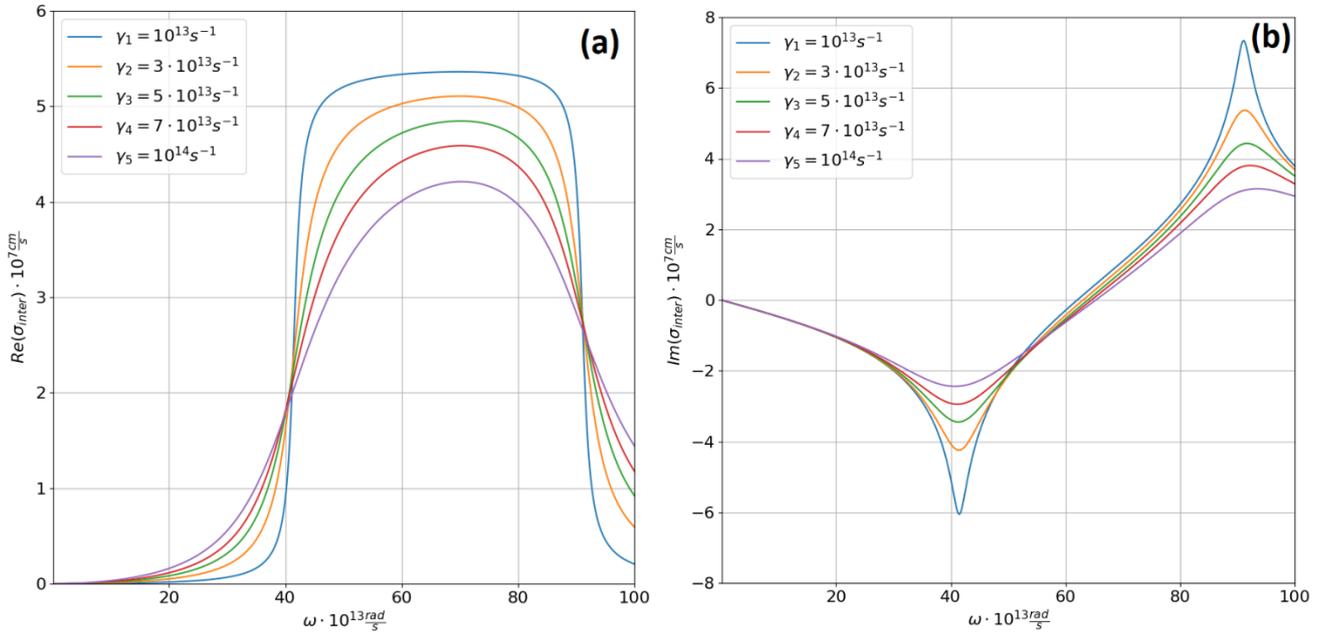

Fig. 4. Numerically calculated real (a) and imaginary (b) parts of the intersubband surface conductivity at $E_f = -0.5\ eV$ and various relaxation rates (see inset).

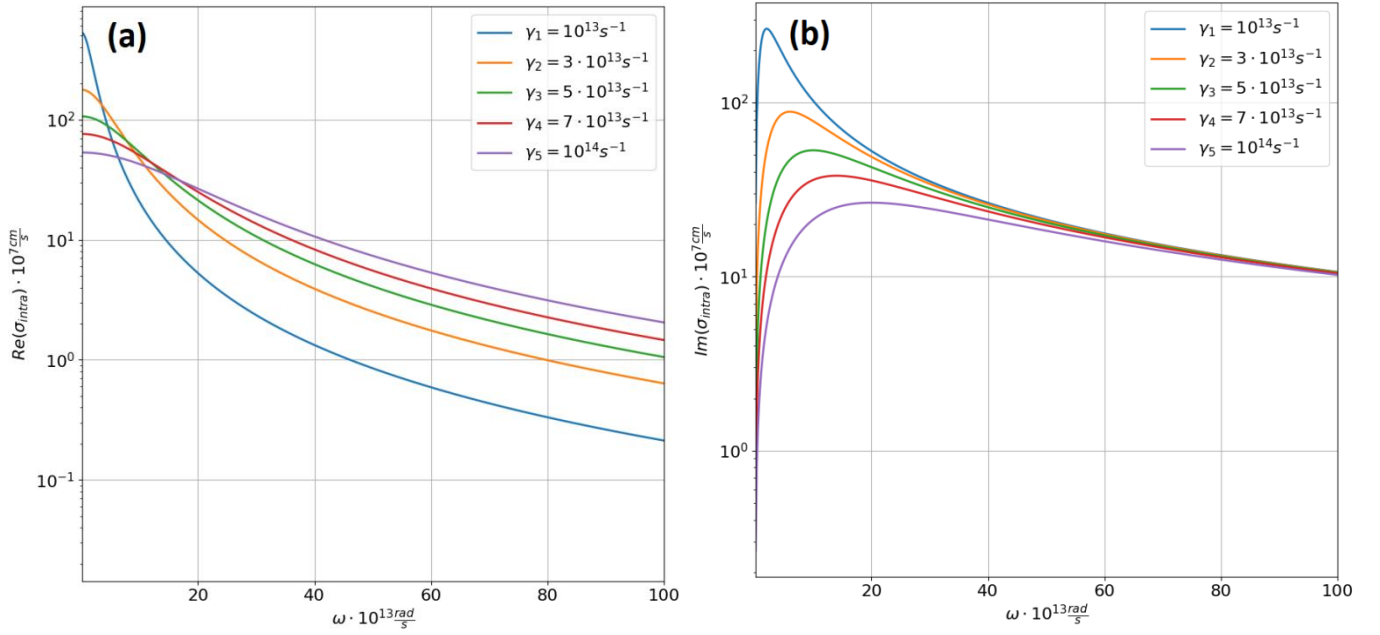

Fig. 5. Numerically calculated real (a) and imaginary (b) parts of the intrasubband conductivity at $E_f = -0.5\ eV$ and various relaxation rates (log-linear axis).

The graphs of the total conductivity are shown in Fig. 6. It is seen from Figures 4, 5 and 6, that the intrasubband part of the surface conductivity is dominant in the far-IR range $\omega = 10^{13}\ ...\ 10^{14}\ \frac{rad}{s}$, while in the range $\omega = 10^{14}\ ...\ 10^{15}\ \frac{rad}{s}$ the intersubband part dominates and the plateau in the frequency interval from $\frac{\Delta E_{min}}{\hbar}$ to $\frac{\Delta E_{max}}{\hbar}$ appears (compare to Fig. 2).

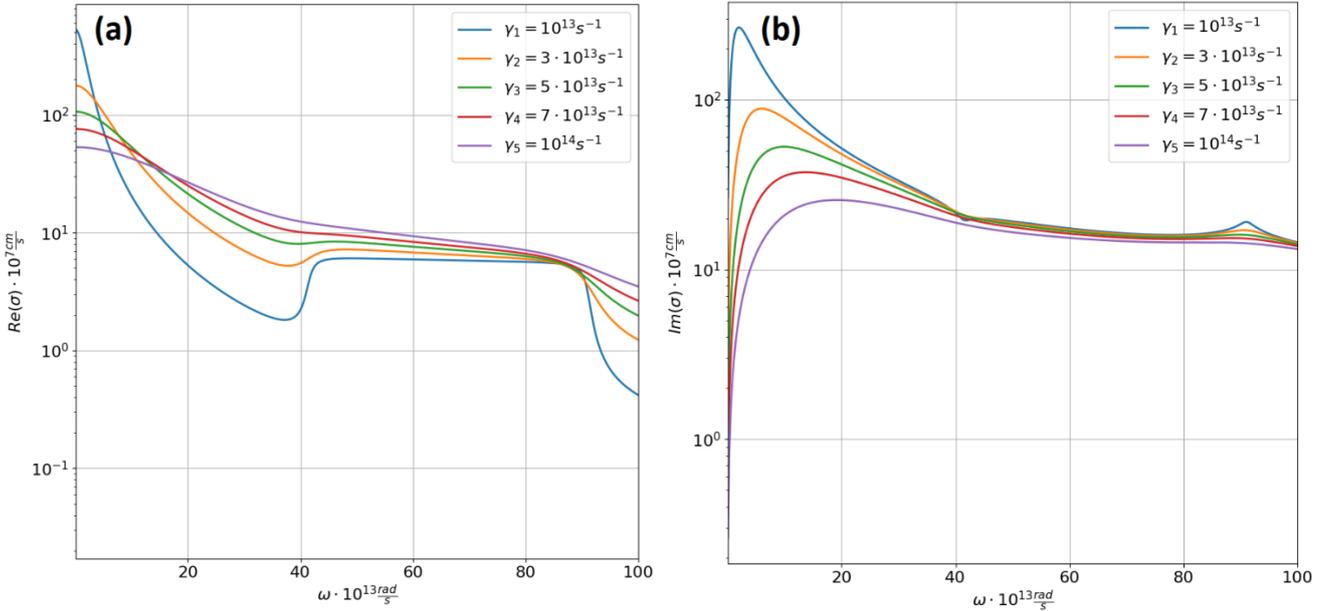

Fig. 6. Numerically calculated real (a) and imaginary (b) parts of the surface conductivity at $E_f = -0.5\ eV$ and various relaxation rates (log-linear axis).

Now consider how the conductivity behavior changes with the change of the Fermi level (at $\gamma = 10^{13} s^{-1}$). As can be seen from Fig. 7, with an increase of the absolute value of the Fermi level, the above-mentioned features of the intersubband absorption become more clear.

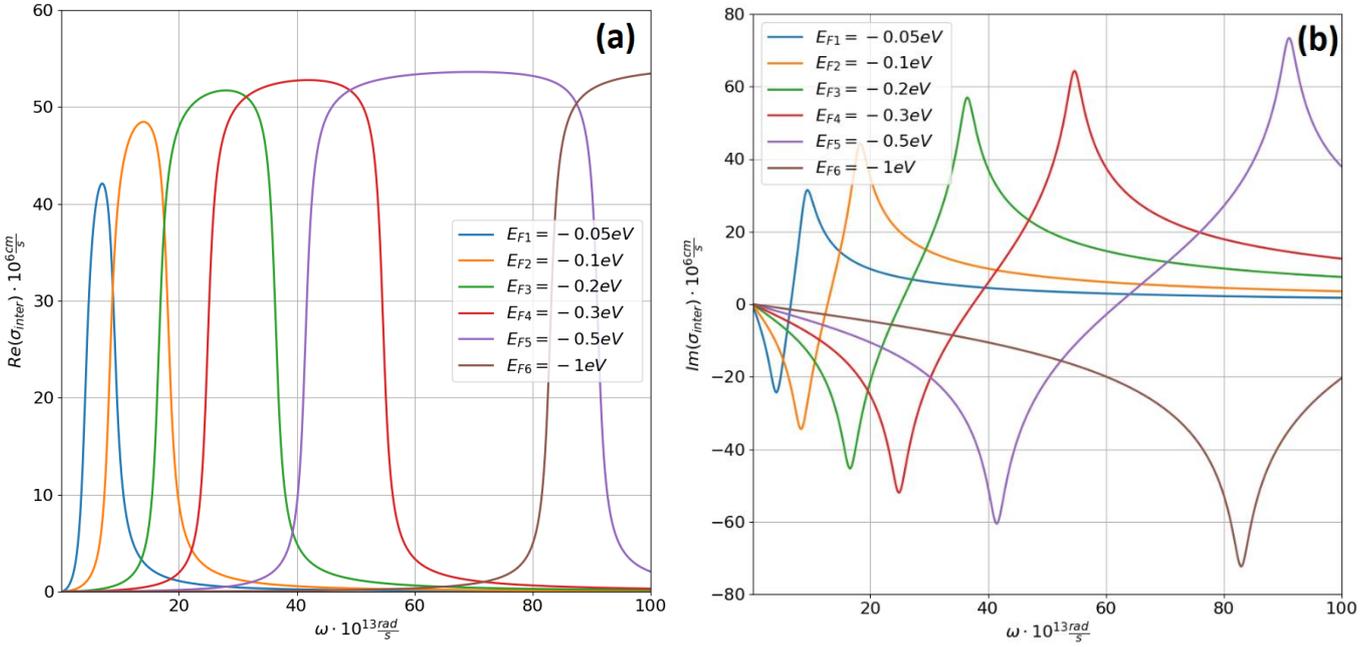

Fig. 7. Numerically calculated real (a) and imaginary (b) parts of the intersubband conductivity at $\gamma = 10^{13} s^{-1}$ and various Fermi levels.

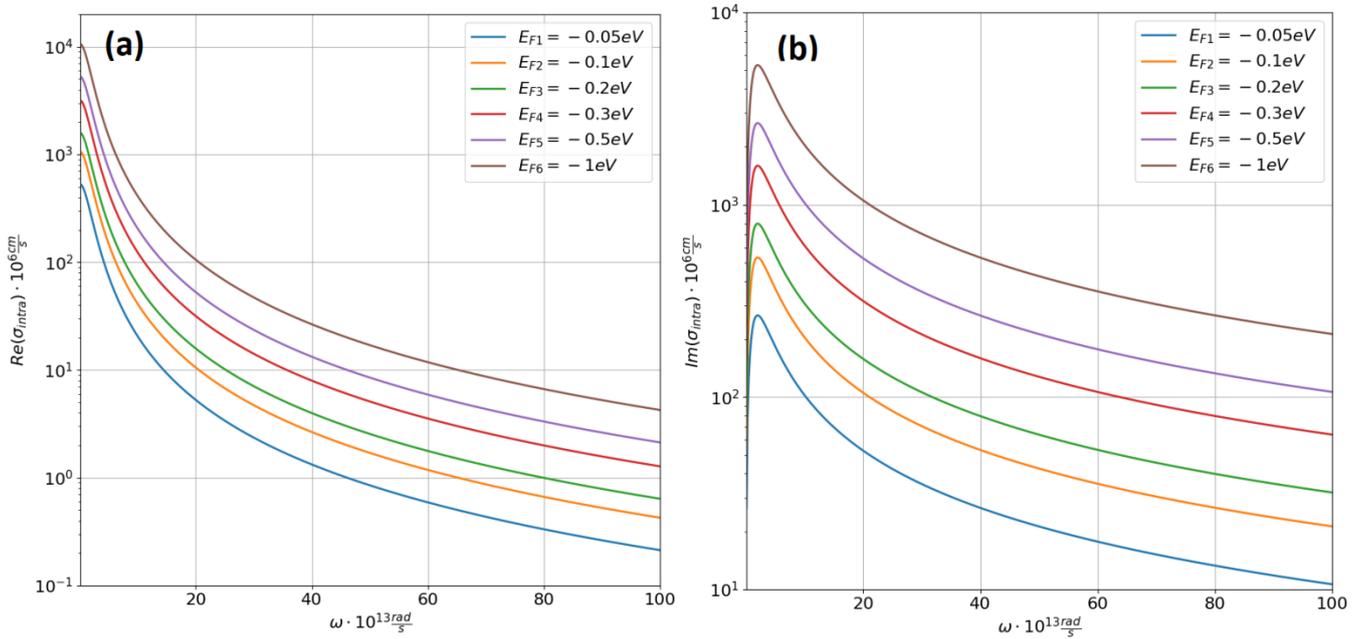

Fig. 8. Numerically calculated real (a) and imaginary (b) parts of the intraband conductivity at $\gamma = 10^{13} s^{-1}$ and various Fermi levels (log-linear axis).

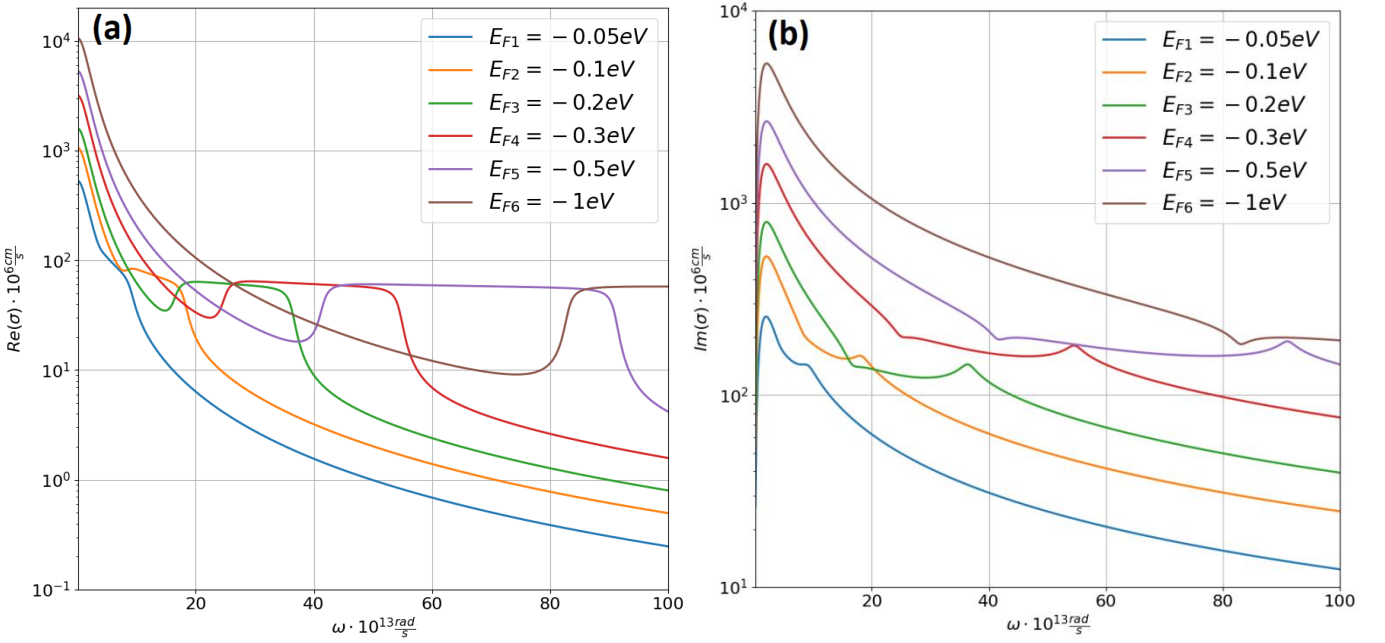

Fig. 9. Numerically calculated real (a) and imaginary (b) parts of the total surface conductivity at $\gamma = 10^{13}\ s^{-1}$ and various Fermi levels (log-linear axis).

*B. Transmission and reflection coefficients*

Figure 10 shows the dependence of the reflection coefficient (in terms of total power of electromagnetic radiation, see expressions in Section IV. A) on Fermi level. Calculations were carried out for graphane on a silicon carbide (SiC) substrate with permittivity $\epsilon = 9.7$ in IR range.

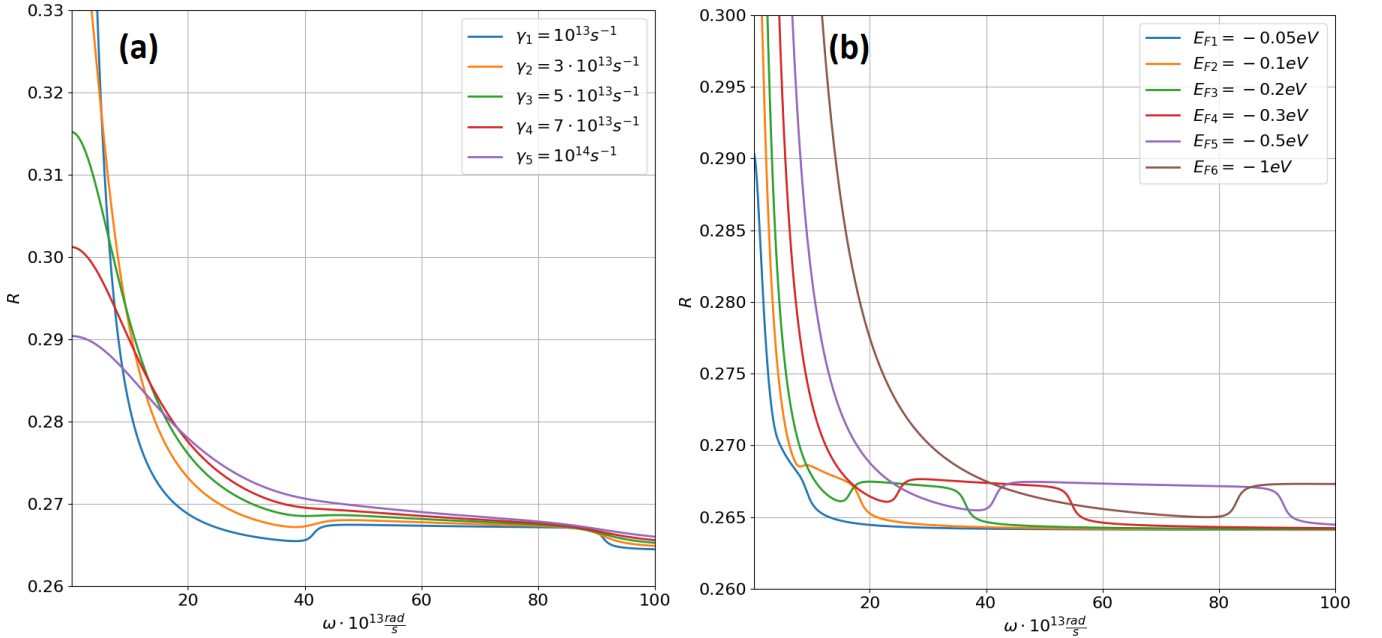

Fig.10. Power reflection coefficient at various relaxation rates for $E_F = -0.5\ eV$ (a) and different Fermi levels for $\gamma = 10^{13} s^{-1}$ (b).

Figure 10 shows that the typical features of intersubband surface conductivity still manifest themselves as plateaus, which become flatter when a relaxation rate increases. With the increase of the absolute value of a Fermi level, plateau widens and shifts toward higher frequencies. That is why measuring of absorption profile makes it possible to determine independently a Fermi level in a graphane sample or, alternatively, to clarify the parameters of the band structure splitting if the Fermi level is already known from some other measurements. Figure 11 shows similar dependencies for the coefficient $W = 1 - R - T$, which characterizes the energy absorption in the graphane sample.

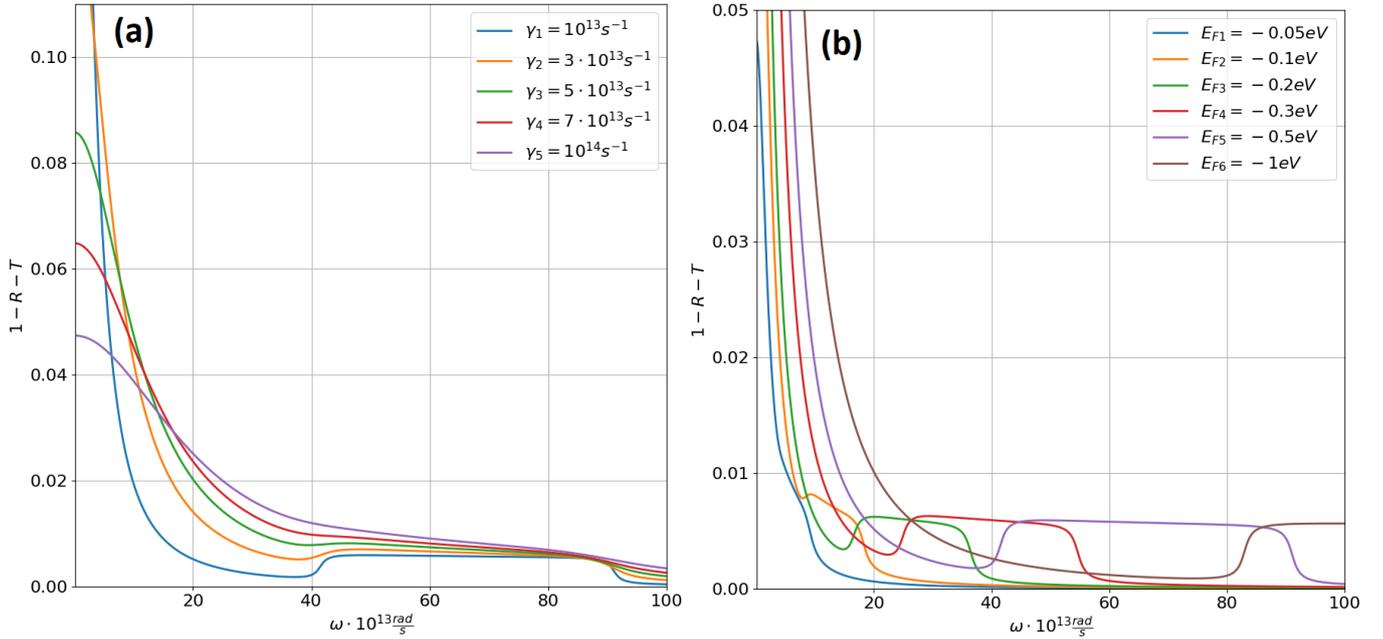

Fig.11. The energy absorption coefficient at various relaxation rates for $E_F = -0.5\ eV$ (a) and at various Fermi levels for $\gamma = 10^{13} s^{-1}$ (b).

*C. Dispersion of 2D plasmon-polaritons*

In this section we consider graphane on a silicon carbide substrate ($\epsilon = 9.7$) and use the expressions obtained in Section IV.B for numerical calculations. In Figure 12 the frequency dependencies of real and imaginary parts of surface plasmon-polariton wavenumber are show for a constant Fermi level $E_F = -0.5\ eV$ and various relaxation rates.

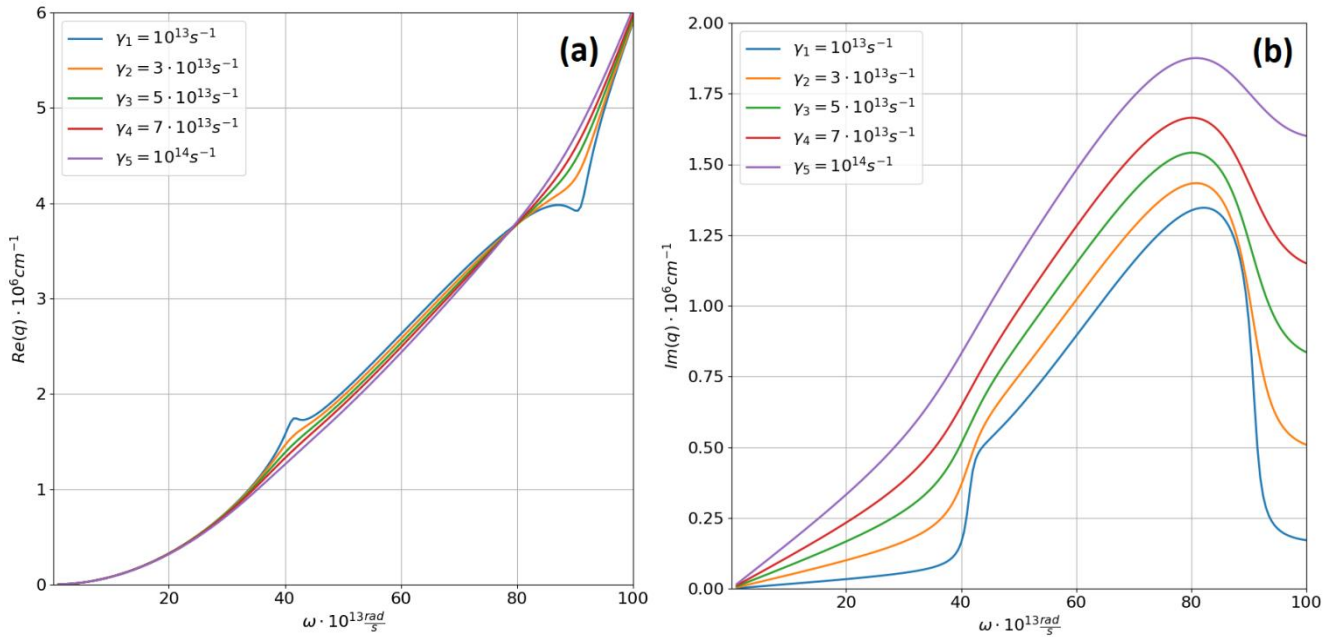

Fig. 12. The real (a) and imaginary (b) parts of surface plasmon-polaritons wavenumbers at different relaxation rates $\gamma = 10^{13} \ldots 10^{14}\ s^{-1}$ and $E_f = -0.5\ eV$.

The graphs for the quality factor $Q = \frac{\text{Re}(q)}{\text{Im}(q)}$ are plotted in Figure 13. This value characterizes the distance of SPP propagation before damping, measured in reversed wavenumbers. As a consequence, one can see a dip on the dependence $Q(\omega)$ associated with the previously mentioned plateau of the interband conductivity. Also there is a local maximum of $Q(\omega)$ at the frequency $3.5 \cdot 10^{14} \frac{rad}{s}$ which is a left boundary of this plateau. The amplitude of this peak almost linearly decreases with the relaxation rate increase.

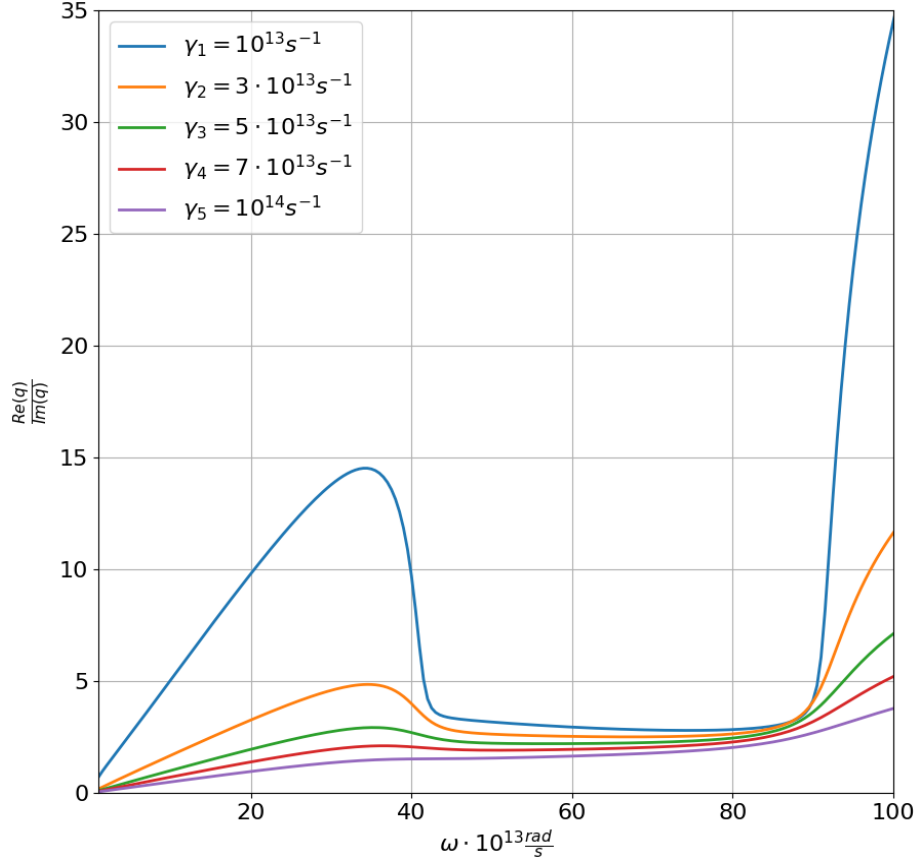

Fig. 13. Quality factor $Q(\omega) = \frac{\text{Re}(q)}{\text{Im}(q)}$ for surface plasmon-polaritons at different relaxation rates $\gamma = 10^{13} \dots 10^{14} \ s^{-1}$ and $E_f = -0.5 \ eV$.

Figures 14 and 15 show the characteristics of surface plasmon-polaritons at the constant relaxation rate $\gamma = 10^{13} s^{-1}$ and different Fermi levels.

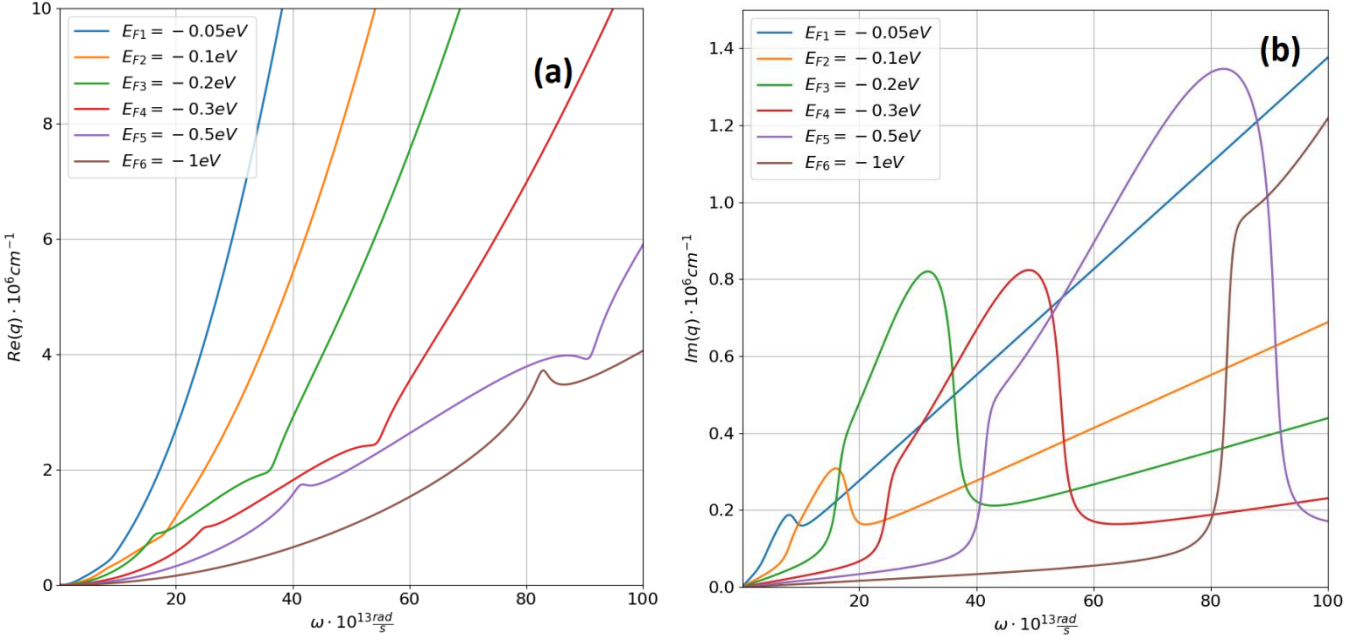

Fig. 14. The real (a) and imaginary (b) parts of the surface plasmon-polariton wavenumber at various $E_F = -0.05 \ldots -1\ eV$ and $\gamma = 10^{13} s^{-1}$.

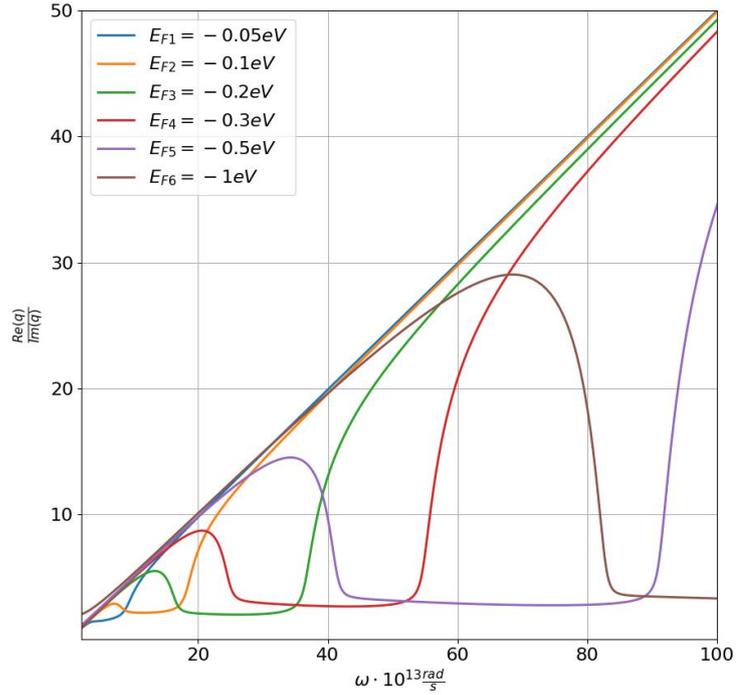

Fig. 15. Quality factor $Q(\omega) = \frac{\text{Re}(q)}{\text{Im}(q)}$ of surface plasmon-polaritons at various $E_F = -0.05 \ldots -1\ eV$ and $\gamma = 10^{13} s^{-1}$.

As a Fermi level increases, the dip on the dependence $Q(\omega)$ shifts toward shorter wavelenghs and becomes wider (see Fig. 15). Outside this area the function $Q(\omega)$ can be approximated as $\frac{\text{Re}(q)}{\text{Im}(q)} \approx \frac{\omega}{2\gamma}$. This approximation is valid in the region where the intrasubband component of complex conductivity dominates. Indeed, assuming the parameter $\epsilon$ in Eq. (15) to be real, we obtain:

$$q = i(1+\epsilon)\frac{\omega}{4\pi\sigma};$$

$$q \propto \frac{i}{\sigma}; \frac{Re(q)}{Im(q)} = \frac{Im(\sigma)}{Re(\sigma)} \approx \frac{Im(\sigma_{intra})}{Re(\sigma_{intra})}.$$

Next, using Eq. (8), we get:
$$\sigma_{intra}(\omega) = C\frac{i}{1+\frac{i2\gamma}{\omega}},$$
where $C = \frac{e^2 g}{4\pi\hbar^2\omega}\sum_{\pm}\left|k\frac{\partial E_{\pm}(k)}{\partial k}\right|_{E_{\pm}=E_F}$ is a real constant, $\gamma_{intra} = \gamma$. Finally, as stated above, we obtain $\frac{Re(q)}{Im(q)} \approx \frac{\omega}{2\gamma}$.

## VI. Summary

In this paper we have developed the theory of graphane optical response in the infrared and terahertz regions. This theory is based on a simple analytical model of graphane band structure in the vicinity of Γ-point and on the modified model of quantum coherence relaxation. Comparison of the results of our calculations with the experimental data seems to be a reliable method for clarification of the used band structure model, proposed in recent literature. Also the results obtained in this work can be used to determine the important parameters of charge carriers in graphane like Fermi level and relaxation rate and, in this way, to estimate the prospects for creating new optical devices based on this novel material.

## Acknowledgements

E. I. Preobrazhensky and I. V. Oladyshkin are grateful to Russian Science Foundation for support (project #21-72-00076, https://rscf.ru/project/21-72-00076/). M.D. Tokman is grateful for support to Russian Foundation for Basic Research (project #20-02-00100 A).

## Appendix A. A presence of a gap between subbands.

The dependences $E_\pm(k^2)$ given by the Hamiltonian (1) correspond to touching subbands at the point $k = 0$. We generalize the results obtained to the case of a finite gap. To describe this situation, we modify the Hamiltonian (1):

$$\hat{H}_k = -a\hat{I}(k_x^2 + k_y^2) - b[\sigma_z(k_x^2 - k_y^2) + 2\sigma_x k_x k_y + \hat{\sigma}_y \mathcal{K}^2] \tag{A1}$$

where $\mathcal{K}$ is the additional constant with the dimension of a wavenumber, $\hat{\sigma}_y$ is the y-Pauli matrix. The Hamiltonian (A1) corresponds to two parabolic-like curves $E_\pm(k^2)$:

$$E_\pm = -ak^2 \pm b\sqrt{k^4 + \mathcal{K}^4}, \tag{A2}$$

which are separated by the gap of the width $W_{vv} = 2b\mathcal{K}^2$ (Fig. A1).

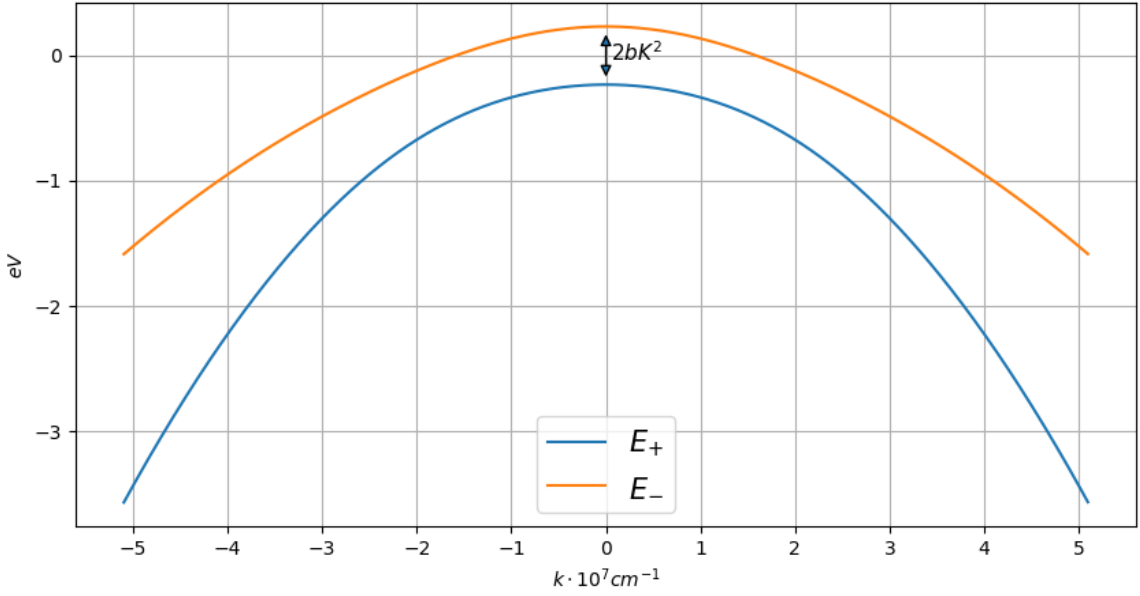

Fig.A1. The model of valence with a gap, $\mathcal{K} = 2.5 \cdot 10^7 cm^{-1}$.

In this case we get:

$$\begin{pmatrix} u_{1k} \\ u_{2k} \end{pmatrix}_\pm = \frac{1}{\sqrt{N_\pm}} \begin{pmatrix} -\frac{2k_x k_y - i\mathcal{K}^2}{k_x^2 - k_y^2 \mp \sqrt{k^4 + \mathcal{K}^4}} \\ 1 \end{pmatrix}, \tag{A3}$$

where $N_\pm = \frac{2\sqrt{k^4+\mathcal{K}^4}}{\sqrt{k^4+\mathcal{K}^4} \mp (k_x^2 - k_y^2)}$;

$$|\mathcal{X}_{(+)k(-)k}|^2 = \frac{1}{k^4+\mathcal{K}^4} \times \frac{\left(k_y(\sqrt{k^4+\mathcal{K}^4} - k^2) - \frac{2k_x^2 k^2 k_y}{\sqrt{k^4+\mathcal{K}^4}}\right)^2 + \mathcal{K}^4 k_x^2 \left(1 + \frac{k^2}{\sqrt{k^4+\mathcal{K}^4}}\right)^2}{\left(k_x^2 - k_y^2 + \sqrt{k^4+\mathcal{K}^4}\right)^2} \tag{A4}$$

It can be proved that in the limit $k^2 \gg \mathcal{K}^2$ the expressions (A3) and (A4) are equal to the expressions (3) and (4), respectively. For $a \sim b$ the strong inequality $k^2 \gg \mathcal{K}^2$ is equivalent to the condition: $E_{(+)}(k^2) - E_{(-)}(k^2) \gg W_{vv}$.

In the expressions (8) and (9), describing the conductivity components, there is the following ratio for the value $\Delta E \left|\frac{d\Delta E}{dk}\right|^{-1}$:

$$\Delta E \left|\frac{d\Delta E}{dk}\right|^{-1} = \frac{k^4 + \mathcal{K}^4}{2k^3}. \tag{A5}$$

In the case $|E_F| < b\mathcal{K}^2 = W_{vv}/2$ there is $\Delta E_{min} = 2b\mathcal{K}^2 = W_{vv}$ in Eq. (9). In the limit of enough large Fermi energy, when $\frac{|E_F|}{a+b} \gg \mathcal{K}^2$ (i.e. $\Delta E_{min} \gg W_{vv}$), both models (i.e. given by the Hamiltonians Eqs. (1) and (A1) respectively) lead to the same results.

Let us consider how surface conductivity and a wavenumber of surface plasmon-polaritons change at different values of the parameter $\mathcal{K}$.

Calculations were carried out with the following parameters: $\mathcal{K}_1 = 0$, $\mathcal{K}_2 = 0.5 \cdot 10^7\ cm^{-1}$, $\mathcal{K}_3 = 10^7\ cm^{-1}$, $\mathcal{K}_4 = 1.6 \cdot 10^7\ cm^{-1}$, $\mathcal{K}_5 = 2 \cdot 10^7\ cm^{-1}$; $\gamma = 5 \cdot 10^{13}\ s^{-1}$, $E_F = -0.1\ eV$ (a Fermi level is counted from the zero energy level corresponding to the arithmetic mean between the maxima of the subbands).[5]

Fig. A2 shows the results of numerical calculations of the real (a) and imaginary (b) parts of the surface conductivity; the inset (c) shows band structures at various parameters $\mathcal{K}$.

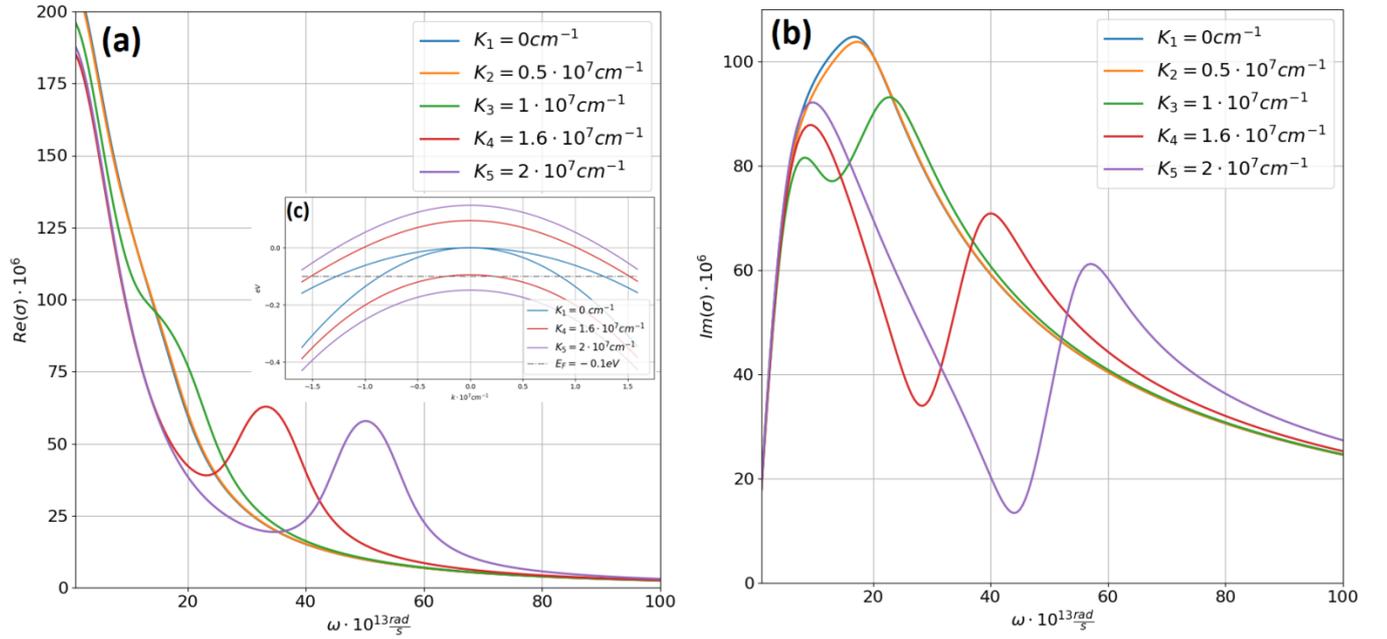

Fig. A2. Numerical calculations of the real (a) and imaginary (b) parts of surface conductivity at various parameters $\mathcal{K}$ and $\gamma = 5 * 10^{13}\ s^{-1}$, $E_f = -0.1\ eV$; box (c) is band structures at various parameters $\mathcal{K}$.

Numerical calculations demonstrate that the real part of surface conductivity dominates at frequencies $10^{13} - 10^{14}\ \frac{rad}{s}$. And as the $\omega$ increases, the imaginary component begins to prevail. The Fig. A3 shows real (a), imaginary (b) parts of the surface plasmon-polariton wavenumber and the quality factor (c) $Q = \frac{Re(q)}{Im(q)}$.

---

[5] If another zero energy level is selected, for example, it is the maximum of the upper subband, the replacement variable should be used $E_F \Rightarrow E_{F\ new} = E_F - |b|\mathcal{K}^2$, or one needs to shift the functions $E_\pm$ by the similar value.

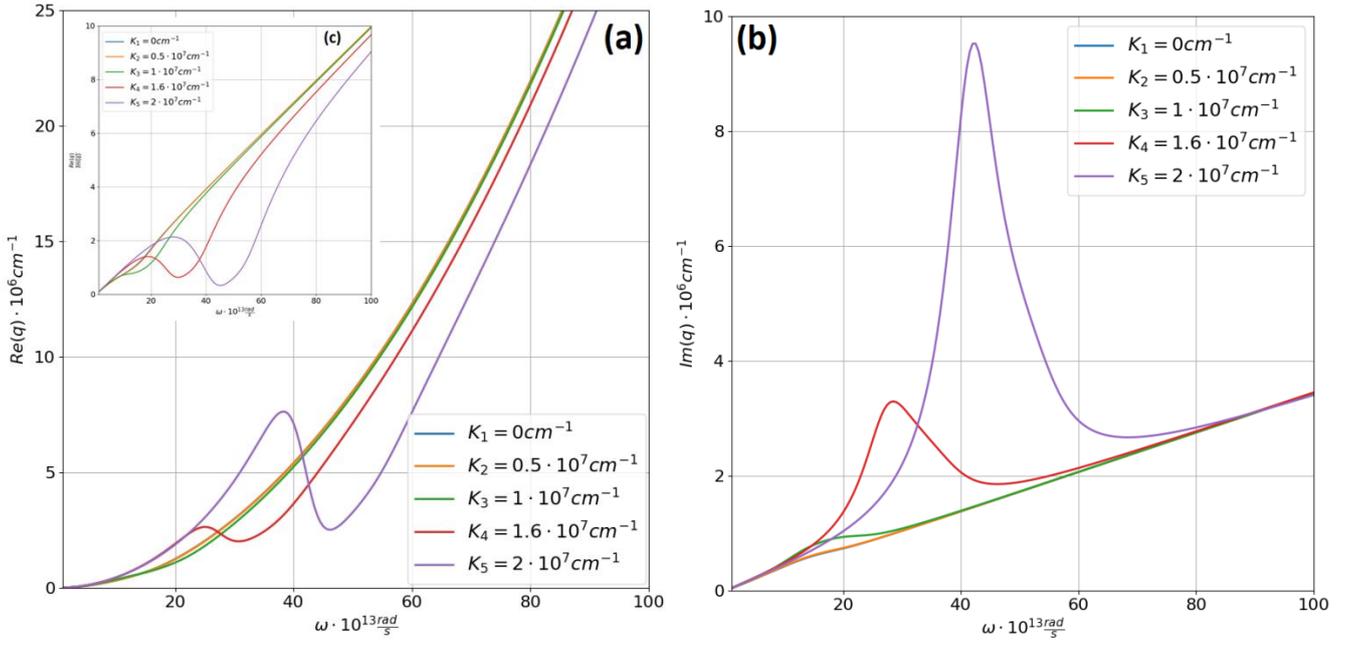

Fig. A3. Real (a) and imaginary (b) parts of the surface plasmon-polaritons wavenumber at $\gamma = 5*10^{13}\ s^{-1}$, $E_f = -0.1\ eV$ and various parameters $\mathcal{K}$; (c) the quality factor $Q = \frac{\text{Re}(q)}{\text{Im}(q)}$.

The presence of the gap between the subbands significantly affects the surface plasmon-polaritons wavenumber. Fig. A3 shows that a characteristic oscillation appear in the real part of plasmon dispersion curve.